\documentclass[10pt, final]{IEEEtran}    
 \pdfoutput=1 
\usepackage{arydshln}
\usepackage{setspace}

\usepackage{multirow}
\usepackage{booktabs}
\usepackage{setspace}

\IEEEoverridecommandlockouts
\usepackage{graphics} 
\usepackage{epsfig} 
\usepackage{times} 

\usepackage{algorithm}
\usepackage[noend]{algpseudocode}

\makeatletter
\def\BState{\State\hskip-\ALG@thistlm}
\makeatother
\usepackage{amssymb}
\usepackage{pifont}
\newcommand{\cmark}{\ding{51}}%
\newcommand{\xmark}{\ding{55}}%

\usepackage{amssymb}

\usepackage{amsthm}
\usepackage{amsmath} 
\usepackage{amssymb}  
\usepackage{epstopdf}
\usepackage{epsfig}
\usepackage{subfigure}
\newcommand{\m}{\boldsymbol}
\usepackage{tikz}
\usetikzlibrary{dsp,fit}
\usetikzlibrary{decorations.pathmorphing} 

\usetikzlibrary{dsp,fit}
\usetikzlibrary{decorations.pathmorphing} 
\tikzstyle{block} = [draw,rectangle,thick,minimum height=3em,minimum width=5em]
\tikzstyle{branch} = [circle,inner sep=1pt,minimum size=2mm,fill=white,draw=black]
\tikzstyle{branch1} = [inner sep=1pt,minimum size=0.1mm,fill=white,draw=black]
\tikzstyle{branch1} = [circle,inner sep=0pt,minimum size=1mm,fill=black,draw=black]

\usepackage{enumitem}

\setlist[itemize]{leftmargin=*}

\makeatletter
\dspdeclareoperator{dspvoidshapeadder}{
    \pgfutil@tempdima=\radius
    \pgfutil@tempdima=0.55\pgfutil@tempdima
    \pgfusepathqstroke
}
\tikzstyle{connector2} = [-,thick]
\tikzstyle{connector} = [->,thick]
\tikzstyle{connector1} = [->,ultra thick]
\tikzstyle{snakeline} = [connector, decorate, decoration={pre length=0.2cm,
                         post length=0.2cm, snake, amplitude=.7mm,
                         segment length=2mm},very thick, red, ->]
                         \tikzstyle{snakeline1} = [connector, decorate, decoration={pre length=0.2cm,
                                                  post length=0.2cm, snake, amplitude=.3mm,
                                                  segment length=2mm},very thick, blue, ->]
  \tikzstyle{snakeline2} = [connector, decorate, decoration={pre length=0.4cm,
                                                  post length=0.4cm, snake, amplitude=1mm,
                                                  segment length=4mm},very thick, red, ->]
\tikzset{
vdspadder/.style={
  shape=dspvoidshapeadder,
   line cap=rect,
  line join=rect,
  line width=\dspblocklinewidth,
  minimum size=\dspoperatordiameter,
  label={185:$+$},
  label={265:$-$}
  },
vadspadder/.style={
  shape=dspvoidshapeadder,
  line cap=rect,
  line join=rect,
  line width=\dspblocklinewidth,
  minimum size=\dspoperatordiameter,
  label=below right:$-$,
  label=above right:$+$
  }
}
\makeatother
\newtheorem{rem}{Remark}

\newtheorem{assumption}{Assumption}
\usepackage{amsmath}
\usepackage{amssymb}
\usepackage{textcase}
\usepackage{soul}
\usepackage{indentfirst}

\newcommand{\mat}[1]{\boldsymbol{#1}}

\newcommand{\normof}[1]{\|#1\|}

\newcommand{\bmat}[1]{\begin{bmatrix} #1 \end{bmatrix}}

\providecommand{\mA}{\ensuremath{\mat{A}}}
\providecommand{\mB}{\ensuremath{\mat{B}}}
\providecommand{\mC}{\ensuremath{\mat{C}}}
\providecommand{\mD}{\ensuremath{\mat{D}}}
\providecommand{\mE}{\ensuremath{\mat{E}}}

\providecommand{\mG}{\ensuremath{\mat{G}}}

\providecommand{\mI}{\ensuremath{\mat{I}}}

\providecommand{\mK}{\ensuremath{\mat{K}}}
\providecommand{\mL}{\ensuremath{\mat{L}}}
\providecommand{\mM}{\ensuremath{\mat{M}}}
\providecommand{\mN}{\ensuremath{\mat{N}}}

\providecommand{\mP}{\ensuremath{\mat{P}}}

\providecommand{\mU}{\ensuremath{\mat{U}}}
\providecommand{\mV}{\ensuremath{\mat{V}}}

\providecommand{\mY}{\ensuremath{\mat{Y}}}

\newcommand{\EQ}{\begin{equation}\begin{array}{lllllllll}}
\newcommand{\EE}{\end{array}\end{equation}}
\newcommand{\MT}{\left[ \begin{array}{ccccccccc}}
\newcommand{\EM}{\end{array}\right]}

\expandafter\def\expandafter\normalsize\expandafter{%
    \normalsize
    \setlength\abovedisplayskip{2.2pt}
    \setlength\belowdisplayskip{2.2pt}
    \setlength\abovedisplayshortskip{2.2pt}
    \setlength\belowdisplayshortskip{2.2pt}
}

\begin{document}

\title{
Dynamic State Estimation under Cyber Attacks: A Comparative Study of Kalman Filters and Observers}
\author{Ahmad F. Taha,~\IEEEmembership{Member,~IEEE, }
        Junjian~Qi,~\IEEEmembership{Member,~IEEE, }
         Jianhui Wang,~\IEEEmembership{Senior Member,~IEEE }
        \\ and Jitesh H. Panchal,~\IEEEmembership{Member,~IEEE}
        \thanks{ 
        
        A. F. Taha is a research intern with the Energy Systems Division, Argonne National Laboratory, Lemont IL 60439 USA in 2015 and a Ph.D. candidate at the School of Electrical and Computer Engineering, Purdue University, West Lafayette, IN 47907 USA  (e-mail: tahaa@purdue.edu).
 
        J. Qi and J. Wang are with the Energy Systems Division, Argonne National Laboratory, Lemont, IL 60439 USA (e-mail: jqi@anl.gov; jianhui.wang@anl.gov).
        
        J. H. Panchal is with the School of Mechanical Engineering, Purdue University, West Lafayette, IN 47907 USA (e-mail: panchal@purdue.edu).
        }
}

\maketitle

\begin{abstract}
Utilizing highly synchronized measurements from synchrophasors, 
dynamic state estimation (DSE) can be applied 
for real-time monitoring of smart grids. 
Concurrent DSE studies for power systems are intolerant to unknown inputs and potential attack vectors --- a research gap we will fill in this work. Particularly, we (a) present an overview of concurrent estimation techniques, highlighting key deficiencies, (b) develop DSE methods based on cubature Kalman filter and dynamic observers, (c) rigorously examine the strengths and weaknesses of the proposed methods under attack-vectors and unknown inputs, and (d) provide comprehensive recommendations for DSE. Numerical results and in-depth remarks are also presented.
\end{abstract}

\vspace*{-0.5cm}

\section{Introduction and Motivation}


\IEEEPARstart{S}{tate} estimation is an important application of the energy management system. 
However, most widely studied static state estimation methods \cite{se1,se2,se3,se5,qi2012power} cannot effectively capture the dynamics of power systems due to its dependency on slow update rates of supervisory control and data acquisition (SCADA) systems. 
In contrast, dynamic state estimation (DSE) enabled by phasor measurement units (PMUs), devices with high update rates and high global positioning system (GPS) synchronization accuracy, 
can provide accurate dynamic states of the system, and will thus play a critical role in achieving real-time wide-area monitoring, protection, and control in smart grids. 
The main objective of North American SynchroPhasor Initiative (NASPI)
is to 
improve the visibility and reliability of power systems by wide area measurement and control
\footnote{available at https://www.naspi.org/.}. Subsequently, efficient DSE methods can be utilized to improve the visibility of power networks.


The main goal of this paper is to 
present DSE alternatives that address major limitations of current DSE methods such as intolerance to unknown inputs, attack-vectors, and parametric uncertainty. 
To achieve that, we study DSE methodologies for power systems, utilizing nonlinear, dynamic observers and recently developed, efficient Kalman filters.  Many of the recent research efforts on DSE heavily focus on Kalman filter-based methods for reasonable state estimation. We show in this paper that this approach has its own limitations, in comparison with observers for nonlinear dynamical systems.

Consequently, our objective is not to develop theory for state estimators or observers for generic dynamic systems, but to (a) survey the literature for current DSE methods, (b) discuss the 
limitations of the current status quo of DSE, (c) present and apply better alternatives, 
and (d) discuss the shortcomings and strengths of the proposed solution.

\section{Literature Review and Paper Contributions}\label{sec:literature} 

Here, we present an overview of DSE methods, introduce dynamic observers, 
and highlight our main contributions.

\subsection{Kalman Filters for Dynamic State Estimation}

Unlike many estimation methods that are either computationally unmanageable or require special assumptions about the form of the process and
observation models, Kalman filter (KF) only utilizes the first two moments of the state (mean and covariance) in its update rule \cite{kalman1960new,ukf1}. 
It consists of two steps: in prediction step, the filter propagates the estimate from last time step to current time step; 
in update step, the filter updates the estimate using collected measurements. 

KF was initially developed for linear systems. However, for DSE the system equations and outputs have strong nonlinearity. 
Thus variants of KF which can deal with nonlinear systems have been introduced. 
For example, DSE has been implemented by extended Kalman filter (EKF) \cite{ekf1}, \cite{ekf2}, unscented Kalman filter (UKF) \cite{pwukf1}, \cite{pwukf2}, \cite{ghahremani2011online}, square-root unscented Kalman filter \cite{qi}, \cite{van2001square}, extended particle filter \cite{zhou}, and ensemble Kalman filter \cite{zhou2015dynamic}. 
The reader is referred to~\cite{Simon2006} for a survey of DSE techniques.

\subsection{Observers for Dynamical Systems and Power Networks}

Dynamic observers have been thoroughly 
investigated for different classes of systems. To mention a few, they have been developed for: linear time-invariant~(LTI) systems, nonlinear time-invariant (NLTI) systems, LTI and NLTI systems with unknown inputs, sensor and actuator faults, stochastic dynamical systems, and linear and nonlinear hybrid systems. 

Most 
observers 
 utilize the plant's outputs and inputs to generate real-time estimates of the plant states, unknown inputs, and sensor faults. 
 The cornerstone 
 is the innovation function --- sometimes a simple gain matrix designed to nullify the effect of unknown inputs and faults. Linear and nonlinear functional observers, sliding-mode observers, unknown input observers, and observers for fault detection and isolation are all examples on developed observers for different classes of systems, under different assumptions. 

In comparison with KF techniques, observers have not been fully utilized in DSE. 
However, they inherently possess the computational, technical, and theoretical capabilities to perform good estimation of the grid's physical status. 


\subsection{Differences between Kalman Filters and Observers}


Observers are typically used for deterministic systems, whereas KFs are used for stochastic dynamical systems. 
If statistical information on process and measurement noise is available, KFs can be utilized by assuming a certain distribution of noise. 
However, this assumption can be very strict as quantifying distributions of noise is challenging.

While some deterministic observers may be robust and tolerant against process and measurement noise, 
KFs are inherently designed to deal with model approximations and measurement noise. 
As for implementation, observers are simpler than KFs. For observers, matrix gains are computed offline to guarantee the asymptotic stability of the estimation error.

\subsection{Paper Contributions}

The contributions of this paper are summarized as follows.
\begin{enumerate}
	\item Surveying the most widely used DSE methods and discussing their shortcomings;
	\item Extending existing DSE methods by introducing Cubature Kalman filter (CKF), which directly uses nonlinear model as UKF but has better numerical stability; 
	\item Presenting two dynamic observers for nonlinear systems and applying them to power systems; and
	\item Exhaustively studying the experimental and theoretical limitations and strengths of the proposed estimation methods under different adverse scenarios.
\end{enumerate}

The paper is organized as follows. In Section~\ref{sec:SMIBDynamics}, we reproduce the nonlinear dynamics of the single-machine infinite bus system.
The CKF and two dynamic observers are introduced in Sections~\ref{sec:CKF1} and \ref{sec:TwoObservers}. 
Then, numerical results are presented in Section~\ref{sec:simulations}. Finally, insightful remarks, recommendations, and 
conclusions are presented in Sections~\ref{sec:remarks} and~\ref{sec:conc}.

\section{Single-Machine Infinite-Bus System Model}~\label{sec:SMIBDynamics}

Here, we introduce the dynamics of the single-machine infinite bus (SMIB) power system, which
has been used as a benchmark in DSE literature, such as \cite{ekf2,Trip2010,Kamwa2015}.
Let $\m x=\bmat{x_1 & x_2 & x_3 & x_4}^{\top} = \bmat{\delta & \omega & e_q^{\prime} & e_d^{\prime} }^{\top}$ be the internal state of the system, where $\delta$ is the rotor angle ~(rad); $\omega$ is the rotor speed (rad/s); $e_q^{\prime}$ and $e_d^{\prime}$ are the transient voltage along $q$ and $d$ axes. Also, let $\m u = \bmat{u_1 & u_2}^{\top} = \bmat{T_m & E_{f_{d}}}^{\top}$ be the input to the SMIB system, where $T_m$ is the mechanical torque~(p.u.) and $E_{f_{d}}$ is the internal field voltage (p.u.).  The outputs are the real and imaginary part of the voltage phasor and current phasor obtained from PMUs installed at the terminal bus of the generator. We define the output as: $\m y = \bmat{y_1 & y_2 & y_3 & y_4}^{\top} = \bmat{e_R & e_I & i_R & i_I}^{\top}.$
The fast sub-transient dynamics and saturation effects are ignored and the generator model is described by the fourth-order differential equations in the local $d$-$q$ reference frame, as follows:
\begin{align} \label{gen model}
\left\{
    \arraycolsep=1.4pt\def\arraystretch{1.2}
    \begin{array}{l}
      \dot{x}_1=x_2-\omega_0 \\ 
      \dot{x}_2=\frac{\omega_0}{2H}\big(u_1-T_{e}-\frac{K_{D}}{\omega_0}(x_2-\omega_0)\big) \\
      \dot{x}_3=\frac{1}{T'_{d0}}\big(u_2-x_3-(x_{d}-x'_{d})i_{d}\big) \\
      \dot{x}_4=\frac{1}{T'_{q0}}\big(-x_4+(x_{q}-x'_{q})i_{q}\big)
    \end{array}
  \right.
\end{align}
where $T_{e}$ is the electric air-gap torque; $\omega_0$ is the rated value of rotor speed; $H$ and $K_{D}$ are the inertia constant and the damping factor; $i_{q}$ and $i_{d}$ are stator currents; $T'_{q0}$ and $T'_{d0}$ are open-circuit time constants; $x_{q}$ and $x_{d}$ are synchronous reactance; $x'_{q}$ and $x'_{d}$ are transient reactance at the $q$ and $d$ axes, respectively. The $i_{q}$, $i_{d}$, and $T_{e}$ variables in (\ref{gen model}), and the outputs $y_3$ and $y_4$ can be written as functions of $\boldsymbol{x}$ and $\boldsymbol{u}$:
\begin{align*}
\Psi_{R}& = x_4\sin x_1 + x_3 \cos x_1 \\
\Psi_{I}& = x_3\sin x_1 - x_4 \cos x_1 \\
\Psi&=\Psi_R + j\Psi_I \\
I_{t}&=\bar{y}(\Psi - e'_{qI}) \\
y_3&=i_{R}= \operatorname{Re}(I_{t})  \\ 
y_4&=i_{I}= \operatorname{Im}(I_{t})  \\
i_{q}&=\frac{S_B}{S_N}(i_{I}\sin x_1+i_{R}\cos x_1) \\
i_{d}&= \frac{S_B}{S_N}(i_{R}\sin x_1-i_{I}\cos x_1)  \\
e_{q}&=x_3-x'_{d}\,i_{d} \\
e_{d}&=x_4+x'_{q}\,i_{q} \\
T_{e}& \cong P_{e}=\frac{S_B}{S_N}(e_{q}i_{q}+e_{d}i_{d})
\end{align*}
where $\Psi$ is the generator voltage source, $e'_{qI}$ is the transient voltage along $q$ axis of the infinite bus, $\bar{y}$ is the line admittance of the reduced network which only has one line connecting the generator and the infinite bus, and $S_B$ and $S_N$ are the system base MVA and generator base MVA, respectively. The outputs $e_{R}$ and $e_{I}$ can also be obtained as function of $\boldsymbol{x}$ and $\boldsymbol{u}$:
\begin{align}
y_1=e_{R}& = x_4\sin x_1 + e_{q}\cos x_1 \\
y_2=e_{I}& = x_3 \sin x_1 - e_{d}\cos x_1.
\end{align}
Without unknown inputs, actuator faults, or disturbances, the state dynamics can be written in compact form as:
\begin{eqnarray}
\dot{\m x}(t)=\m f(\m x,\m u) \nonumber = \mA \m x(t) + \mB \m u(t) + \boldsymbol\phi(\m x,\m u),~\label{equ:SSnoUI}
\end{eqnarray}
where
\EQ
\label{ABphi}
\small
\m A=
\MT 0 & 1 &  0 & 0 \\
	0 & -\dfrac{K_D}{2H} &  0 & 0 \\
	0 & 0 & -\dfrac{1}{T_{d0}^{\prime}}   & 0 \\
	0 & 0 &  0 & -\dfrac{1}{T_{q0}^{\prime}}   \EM, \notag
\m B=
\MT 0 & 0 \\
	\dfrac{\omega_0}{2H} & 0  \\
	0 & \dfrac{1}{T_{d0}^{\prime}}  \\
	0 & 0  \EM, \notag	\\
\small
\boldsymbol\phi(\m x,\m u)=	
\MT -\omega_0 \\
	\dfrac{\omega_0}{2H} \big(-T_e(\m x,\m u) + K_D \big)	\\
	- \dfrac{x_d-x_d^{\prime}}{T_{d0}^{\prime}}\,i_d(\m x,\m u)  	\\
	\dfrac{x_q-x_q^{\prime}}{T_{q0}^{\prime}}\,i_q(\m x,\m u)   \EM. \notag	
\EE
We utilize the above form of the power system dynamics in the estimation problem in the next sections. We also consider unknown inputs in the system dynamics and attack-vectors against output measurements. 
\begin{rem}
	Albeit only the SMIB system is presented, the estimation methods in this paper can be applied to other systems. 
	More details on the model for multi-machine system can be found in \cite{qi}. 
	We use this simple system as it is the benchmark in many recent DSE methods as in~\cite{Trip2010} and~\cite{Kamwa2015}. 
\end{rem}

\section{Kalman Filter Techniques}~\label{sec:CKF1}

As mentioned above, the current status quo of DSE is based on different variants of KF that can deal with nonlinear models, among which EKF, UKF, and especially CKF are discussed in this section.

\subsection{Extended Kalman Filter}
Although EKF maintains the elegant and computationally efficient recursive update form of KF, it works well only in a `mild' nonlinear environment,  
owing to the first-order Taylor series approximation for nonlinear functions \cite{CKF}. 
It is sub-optimal and can easily lead to divergence.
Also, the linearization can be applied only if the Jacobian matrix exists 
and calculating Jacobian matrices can be difficult and error-prone.
For DSE, EKF has been discussed in \cite{ekf1} and \cite{ekf2}.

\subsection{Unscented Kalman filter}

The unscented transformation (UT) \cite{ut} is developed to address the deficiencies of linearization 
by providing a more direct and explicit mechanism for transforming mean and covariance information. 
Based on UT, Julier et al. \cite{ukf}, \cite{ukf1} propose the UKF as a derivative-free alternative to EKF. 
The Gaussian distribution is represented by a set of deterministically chosen sample points called sigma points. 
The UKF has been applied to power system DSE, for which no linearization or calculation of Jacobian matrices is needed \cite{pwukf1}, \cite{pwukf2}.

However, for the sigma-points, the stem at the center (the mean) is highly significant as it carries more weight which is usually negative for high-dimensional systems. Therefore, in terms of the numerical instability, the UKF is supposed to encounter troubles when used in high-dimensional problems. 

\subsection{Cubature Kalman filter}~\label{sec:CKFF}

Both EKF and UKF can suffer from the curse of dimensionality and the effect of dimensionality may become detrimental in high-dimensional state-space models
especially when there are high degree of nonlinearities in the equations that describe the state-space model \cite{CKF}, \cite{bellman}. 
Making use of the spherical-radial cubature rule, Arasaratnam et al. \cite{CKF} propose CKF, 
which possesses an important virtue of mathematical rigor rooted in the third-degree spherical-radial cubature rule for numerically computing Gaussian-weighted integrals, 
better addressing the numerical instability problem in UKF.

A nonlinear system (without unknown inputs or attack-vectors) can be written in discrete-time form as
\begin{align*} 
     \m{x}_{k}&=\m{f}(\m{x}_{k-1},\m{u}_{k-1})+\m{q}_{k-1} \\
     \m{y}_{k}&=\m{h}(\m{x}_k,\m{u}_k)+\m{r}_k
\end{align*}
where $\m{x}_k \in \mathbb{R}^n$, $\m{u}_k \in \mathbb{R}^v$, and $\m{y}_k \in \mathbb{R}^p$ are 
states, inputs, and observed measurements at time step $k$; the estimated mean and estimated covariance of the estimation error are $\m{m}$ and $\m{P}$; 
$\m{f}$ and $\m{h}$ are vectors consisting of nonlinear state transition functions and measurement functions; $\m{q}_{k-1} \sim N(0,\m{Q}_{k-1})$ is 
the Gaussian process noise at time step $k-1$; $\m{r}_k \sim N(0,\m{R}_k)$ is the Gaussian measurement noise at time step $k$; and 
$\m{Q}_{k-1}$ and $\m{R}_k$ are covariance matrices of $\m{q}_{k-1}$ and $\m{r}_k$.

The procedure of CKF consists of a prediction step and an update step, which are summarized in Algorithms~\ref{algoKF1} and \ref{algoKF2}.  

\begin{algorithm}[!h]
	\caption{CKF Algorithm: Prediction Steps}\label{algoKF1}
	\begin{algorithmic}[1]
		\State \textbf{draw} cubature points from the intersections of the $n$ dimensional unit sphere and the Cartesian axes. 
		\begin{displaymath}
		\m{\xi}_i = \left\{ \begin{array}{ll}
		\sqrt{n}\,e_i, & i=1,\cdots,n \\
		-\sqrt{n}\,e_{i-n}, & i=n+1,\cdots,2n
		\end{array} \right.
		\end{displaymath}
		where $e_i$ is a vector with dimension $n$, whose $i$th element is one and the other elements are zero.
		\State \textbf{propagate} the cubature points. The matrix square root is the lower triangular cholesky factor
		\begin{equation*}
		\m{X}_{i,k-1|k-1}=\sqrt{\m{P}_{k-1|k-1}} \, \m{\xi}_i + \m{m}_{k-1|k-1}.
		\end{equation*}
		\State \textbf{evaluate} the cubature points with dynamic model function
		\begin{equation*}
		\m{X}_{i,k|k-1}^*=\m{f}(\m{X}_{i,k-1|k-1}).
		\end{equation*}
		\State \textbf{estimate} the predicted state mean
		\begin{equation*}
		\m{m}_{k|k-1}=\frac{1}{2n} \sum\limits_{i=1}^{2n}\m{X}_{i,k|k-1}^*.
		\end{equation*}
		\State \textbf{estimate} the predicted error covariance
		\begin{align*}
		\m{P}_{k|k-1}=&\frac{1}{2n}\sum\limits_{i=1}^{2n}\m{X}_{i,k|k-1}^* \m{X}_{i,k|k-1}^{*T} \notag \\
		&-\m{m}_{k|k-1}\m{m}_{k|k-1}^T + \m{Q}_{k-1}.
		\end{align*}
	\end{algorithmic}
\end{algorithm}

\begin{algorithm}[!h]
	\caption{CKF Algorithm: Update Steps}\label{algoKF2}
	\begin{algorithmic}[1]
		\State \textbf{draw} cubature points from the intersections of the $n$ dimensional unit sphere and the Cartesian axes. 
		\State \textbf{propagate} the cubature points
		\begin{equation*}
		\m{X}_{i,k|k-1}=\sqrt{\m{P}_{k|k-1}}\,\m{\xi}_i + \m{m}_{k|k-1}.
		\end{equation*}
		\State \textbf{evaluate} cubature points using measurement function
		\begin{equation*}
		\m{Y}_{i,k|k-1}=\m{h}(\m{X}_{i,k|k-1}).
		\end{equation*}
		\State \textbf{estimate} the predicted measurement
		\begin{equation*}
		\hat{\m{y}}_{k|k-1}=\frac{1}{2n}\sum\limits_{i=1}^{2n}\m{Y}_{i,k|k-1}.
		\end{equation*}
		\State \textbf{estimate} the innovation covariance matrix
		\begin{align*}
		\m{S}_{k|k-1}=&\frac{1}{2n}\sum\limits_{i=1}^{2n}\m{Y}_{i,k|k-1}\m{Y}_{i,k|k-1}^T \notag \\
		&-\hat{\m{y}}_{k|k-1}\hat{\m{y}}_{k|k-1}^T + \m{R}_k.
		\end{align*}
		\State \textbf{estimate} the cross-covariance matrix
		\begin{align*}
		\m{P}_{xy,k|k-1}=&\frac{1}{2n}\sum\limits_{i=1}^{2n}\m{X}_{i,k-1|k-1}\m{Y}_{i,k|k-1}^T \notag \\
		&-\m{m}_{k|k-1}\hat{\m{y}}_{k|k-1}^T.
		\end{align*}
		\State \textbf{calculate} the Kalman gain
		\begin{equation*}
		\m{K}_k=\m{P}_{xy,k|k-1}\m{S}_{k,k-1}^{-1}. 
		\end{equation*}
		\State \textbf{estimate} the updated state
		\begin{equation*}
		\m{m}_{k|k}=\m{m}_{k|k-1}+\m{K}_k(\m{y}_k-\hat{\m{y}}_k).
		\end{equation*}
		\State \textbf{estimate} the updated error covariance
		\begin{equation*}
		\m{P}_{k|k}=\m{P}_{k|k-1} - \m{K}_k\m{P}_{yy,k|k-1}\m{K}_k^T.
		\end{equation*}
	\end{algorithmic}
\end{algorithm}

\section{Utilizing Dynamic Observers for DSE}~\label{sec:TwoObservers}

Here, we compactly present two observers that can be used for DSE. 
First, we develop the system dynamics under state disturbances, actuator faults, unknown inputs and attack vectors. 
Then, the two observers are succinctly presented.

\subsection{Dynamics Under Unknown Inputs and Attack Vectors}
Under disturbances, actuator faults and attack vectors, the dynamics presented in the previous section can be written as
\begin{align}
	\dot{\m x}(t) & = \m f(\m x, \m u)+ \mB_d  \m u_d(t)  + \mB_a  \m f_a(t) ~\label{equ:ssuis}\\
	\m y(t) &= \mC \m x(t)+\m v(t)~\label{equ:ssuivec}
\end{align}
where
$\mB_d$ is the disturbance matrix; $\mB_a$ represents the actuator fault's matrix; $\mC$ denotes the output matrix (see Remark~\ref{rem:output}) and $\m v(t)$ exemplifies a potential attack vector against PMU measurements (see Remarks~\ref{rem:CAs} and~\ref{rem:UIss}).
\begin{rem}~\label{rem:output}
	For simplicity, the output function is assumed to be linear in terms of the internal state of the system. 
	Numerical results indicate that the differences between the outputs from nonlinear and linear output equations, given the same initial conditions and control input, of the SMIB system are minor. 
\end{rem}
\begin{rem}\label{rem:CAs}
	The National Electric Sector Cybersecurity Organization Resource (NESCOR) developed cyber-security failure scenarios with corresponding impact analyses~\cite{NESCOR2014}. 
	The report classifies potential failure scenarios into different classes, including wide area monitoring, protection, and control (WAMPAC) --- the paper's main focus. 
	The following WAMPAC failure scenario motivate the research presented in this paper:
	Measurement Data (from PMUs) Compromised due to PDC Authentication Compromise~\cite{NESCOR2014}. 
	Hence, the addition of attack-vectors against PMU measurements depicts the aforementioned WAMPAC failure scenario.
\end{rem}
\begin{rem}\label{rem:UIss}
	The attack vector, $\m v(t)$, is different from typical measurement noise. While measurement noise vectors are often assumed to follow a certain distribution with small magnitudes, the assumed attack vector here follows no statistical distribution, as demonstrated in the numerical results section. Measurement and process noise are also added. 
\end{rem}
 The unknown inputs to the given system are $\m u_d$~(representing the unknown plant disturbances) and $\m f_a$~(actuator faults). We assume that $\mA, \mB, \mB_d, \mB_a$, and $\mC$ are all known.
 For simplicity, we can combine the unknown inputs $\m u_d$ and $\m f_a$ into one unknown input quantity, $\m d(t)$, defined as follows $\m d(t) = \bmat{\m u_d^{\top}(t) & \m f_a^{\top}(t)}^{\top} \in \mathbb{R}^{d}.$ 
	The matrices $\mB_d$ and $\mB_a$ are combined into $\mD \in \mathbb{R}^{n \times d}$, and the plant dynamics can be rewritten as:
	\begin{eqnarray}~\label{equ:plantDynamics}
	\dot{\m x}(t) &=& \mA \m x(t) + \mB  \m u(t) + \boldsymbol\phi(\m x,\m u)  + \mD  \m d(t)  \\
	\m y(t) &=& \mC \m x(t)+\m v(t). ~\label{equ:plantDynamics2}
	\end{eqnarray}
\begin{rem}\label{rem:UI}
	The exciter output voltage $E_{f_{d}}$ may not available for state-estimation. 
	In that case $E_{f_{d}}$ can be augmented to the unknown input vector and hence, matrices $\mB_1$ and $\mD$ will have different dimensions and values. Ghahremani and Kamwa apply EKF with unknown inputs to generate state estimates 
	where the unknown input is $E_{f_{d}}$~\cite{Kam2011}.
\end{rem}
\begin{assumption}~\label{ass:Obs}
The nonlinear component in SMIB dynamics, $\boldsymbol\phi(\m x, \m u)$, satisfies locally Lipschitz condition in a region $\mathbb{D}$ that includes the origin with respect to $\m x$, uniformly in $\m u$. Precisely, the nonlinear function satisfies the following condition:
$$\normof{\boldsymbol\phi(\m x,\m u) -  \boldsymbol\phi(\m z,\m u) }  \leq  \gamma \normof{\m x - \m z},$$
for all $\m x$ and $\m z$ in $\mathbb{D}$, where $\gamma >0$. If $\mathbb{D}$ is equivalent to $\mathbb{R}^{n}$, the function	$\boldsymbol\phi(\m x, \m u)$ is said to be globally Lipschitz, and it then satisfies the one-sided Lipschitz condition:
$$\langle  \boldsymbol\phi(\m x,\m u)-\boldsymbol\phi(\m z,\m u),\m x - \m z\rangle \leq \rho \normof{\m x - \m z}^2, $$	
where $\rho \in \mathbb{R}$.
\end{assumption}
\begin{rem}
Although we define $\m v(t)$ to be an attack vector, this definition is not restricting. The unknown quantity $\m v(t)$ reflects possible measurement noise or falsely reported measurements. For example, It has been reported that PMUs from multiple vendors might produce conflicting measurements, as highlighted in a North American Synchrophasor Initiative (NASPI) report~\cite{NASPI}. Furthermore, the term $\mD\m d(t)$ models a general class of unknown inputs such as: nonlinearities, modeling uncertainties, noise, parameter variations, unmeasurable system inputs, model reduction errors and actuator faults~\cite{Chen2012,Pertew2005}. 
\end{rem}

\subsection{Observer Design Using Linear Matrix Inequalities (LMI)}~\label{sec:TwoObserversA}

Here, we introduce an observer design for a specific class of nonlinear systems with unknown inputs, which is based on the methods in~\cite{Chen2006} by Chen and Saif. The observer design only assumes that $\mD$ is full-column rank and that the nonlinear function in the state dynamics satisfies Assumption~\ref{algo}. The observer design problem is then formulated as an LMI. 

Define $\hat{\m x}$ to be a real-time estimate of the real state of the system and $\m z$ to be an intermediate state of the estimator. The dynamics of the observer developed in~\cite{Chen2006} can be written as
\begin{eqnarray}
\dot{\m z}(t) & = & \mN \m z(t) + \mG \m u(t) + \mL y(t) + \mM \boldsymbol\phi(\hat{\m x}, \m u) \label{equ:Observer1Dynamics2}\\
\hat{\m x}(t) & = & \m z(t) - \mE \m y(t) \label{equ:Observer1Dynamics}
\end{eqnarray}
where matrices $\mE$ and $\mK$ are design parameters and $\mN, \mG, \mL$ and $\mM$ are obtained from the matrix equalities while guaranteeing the convergence of the state estimates. Algorithm~1 is applied to compute the aforementioned matrices.
\begin{algorithm}[!h]
	\caption{Observer with Unknown Input Design Algorithm}\label{algo}
	\begin{algorithmic}[1]
		\State \textbf{compute} matrices $\mU, \mV, \bar{\m A}$ and $\bar{\m B}$:
		\begin{align*}
		\mU &= -\mD(\mC\mD)^{\dagger}  &	\mV & = \mI - (\mC\mD)(\mC\mD)^{\dagger} \\
		 \bar{\mA}&=  (\mI+\mU\mC)\mA  & \bar{\m B}  &= \mV\mC\mA 
			\end{align*}
			\State \textbf{find} matrices $\bar{\mY}, \bar{\mK}$ and a symmetric positive definite matrix $\mP$ that are a solution for this LMI:
		\begin{equation}~\label{equ:LMIObs1}
		 \bmat{
		 	\boldsymbol\Psi_{11} & \boldsymbol\Psi_{12} \\
		 	\boldsymbol\Psi_{12}^{\top} & \mI_{2n}  
		 } < 0
		\end{equation}
			where $$
			\boldsymbol\Psi_{11} = \bar{\mA}^{\top}\mP + \mP\bar{\mA} + \bar{\mB}^{\top}\bar{\mY}^{\top}  \bar{\mY}\bar{\mB} -\mC^{\top}\bar{\mK}^{\top}-\bar{\mK}\mC+\gamma\mI,$$
			$$	\boldsymbol\Psi_{12}=		\sqrt{\gamma}\left(\mP(\mI+\mU\mC)+\bar{\mY}(\mV\mC)\right)$$
			\State \textbf{obtain} matrices $\mY$ and $\mK$ and the observer parameters $\mN, \mG, \mL$ and $\mM$:
				\begin{eqnarray}
				\mY& = &\mP^{-1}\bar{\mY} \,\, , \,\, \mK=\mP^{-1}\bar{\mK} \nonumber \\
				\mE & = & \mU+\mY\mV ,\,\, \mM \,\, = \,\, \mI +\mE\mC \nonumber \\
				\mN & = & \mM\mA-\mK\mC, \,\,\, \mG \, \, = \,\, \mM\mB,  \,\,\, \label{equ:obsparameters} \\
				\mL & = & \mK (\mI + \mC\mE) -\mM\mA\mE \nonumber
				\end{eqnarray}
	\end{algorithmic}
\end{algorithm}
Given $\mA,\mB,\mC,\mD$ and $\gamma$, the above algorithm finds a solution to the observer dynamics by solving the formulated LMI~\eqref{equ:LMIObs1}. The designed matrix parameters~\eqref{equ:obsparameters} guarantees that the observer dynamics~\eqref{equ:Observer1Dynamics2}--\eqref{equ:Observer1Dynamics} yields asymptotically stable estimation error.

\subsection{Observer Design for Nonlinear Systems}~\label{sec:TwoObserversB} 

Here, we present a recently developed observer in~\cite{Zhang2012} that can be applied for DSE. 
The examined observer assumes the nonlinear function $\boldsymbol\phi(\m x,\m u)$ satisfies Assumption~1. In addition to this assumption, we assume that the nonlinear function is quadratically inner-bounded. Precisely, 
$$\left( \boldsymbol\phi(\m x,\m u)-\boldsymbol\phi(\m z,\m u)\right)^{\top} \left( \boldsymbol\phi(\m x,\m u)-\boldsymbol\phi(\m z,\m u)\right) \leq \mu \normof{\m x -\m z}^{2} $$ $$+ \varphi \,\langle  \boldsymbol\phi(\m x,\m u)-\boldsymbol\phi(\m z,\m u),\m x - \m z\rangle, $$
where $\mu$ and $\varphi$ are real numbers. Similar results related to the dynamics of multi-machine power systems established a similar quadratic bound on the nonlinear component (see.~\cite{Siljak2002}).

The dynamics of the observer can be written as follows:
\begin{equation}~\label{equ:ObserverTwo}
\dot{\hat{\m x}}(t)=\mA\hat{\m x}(t)+\mB\m u(t)+\boldsymbol\phi(\hat{\m x},\m u)+\mL(\m y(t)-\mC\hat{\m x}(t)),
\end{equation}
where $\mL$ is a matrix gain determined as follows. First, given the Lipschitz constants\footnote{Those constants can be determined by evaluating the bounds on the nonlinear function $\boldsymbol\phi(\m x,\m u)$ and adding a possible bound on the unknown inputs and disturbances. Determining those constants affects the design of the observer, and hence it is advised to choose conservative bound-constants on the nonlinear function. The nonlinearities present in the SMIB system are bounded (e.g., sines and cosines of angles, multiplications of bounded quantities such as voltages and currents)} $\rho,\,\varphi$, and $\mu$, the following linear matrix inequality is solved for positive constants $\epsilon_1, \epsilon_2$ and $\sigma$ and a symmetric positive semi-definite matrix $\mP$,
\begin{equation}~\label{equ:LMIObserverDes}
\left[
\begin{array}{c;{2pt/2pt}c}
\mA^{\top}\mP+\mP\mA+(\epsilon_1\rho+\epsilon_2\mu)\mI_n \\-\sigma\mC^{\top}\mC & \mP+\dfrac{\varphi\epsilon_2-\epsilon_1}{2}\mI_n \\\hdashline[2pt/2pt] 
 \left(\mP+\dfrac{\varphi\epsilon_2-\epsilon_1}{2}\mI_n\right)^{\top} & -\epsilon_2\mI_n
\end{array}
\right] 
< 0.
\end{equation}
Then, the observer gain-matrix $\mL$ is given by: 
$$\mL=\frac{\sigma}{2}\mP^{-1}\mC^{\top}.$$
Utilizing the solution to the LMI~\eqref{equ:LMIObserverDes}, the state estimates generated from~\eqref{equ:ObserverTwo} are guaranteed to converge to the actual values of the states.
%
%
%
%
%
%
%
%
%
\section{Numerical Results}~\label{sec:simulations}
In this section, we test the proposed estimation methodologies on the SMIB system under different conditions.

\subsection{System Description and Parameters}~\label{sec:sim22}

\begin{figure}[h]
	\centering
	\includegraphics[scale=0.05]{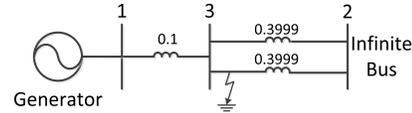}
	\caption{Single machine infinite bus power system.}
	\label{SMIB}
\end{figure}

As shown in Fig. \ref{SMIB}, the SMIB system is extracted from PST toolbox~\cite{chow1992toolbox}. 
We apply a three-phase fault at bus $3$ of line $3-2$ (one of the parallel lines) to generate dynamic response. 
The fault is cleared at near and remote ends after $0.05s$ and $0.1s$. 
DSE is performed for the post-contingency system, for which the parameters can be found in \cite{chow1992toolbox} for the SMIB system.

This system has two known inputs ($T_m$ and $E_{f_{d}}$) and four outputs ($e_R, e_I, i_R,$ and $i_I$). The two known inputs are kept unchanged. Without loss of generality, a system input (such as $E_{f_{d}}$) can be augmented to the unknown input vector and considered as an \textit{unknown quantity} (see Remark~\ref{rem:UI}), in case it is not available. The SMIB's and observer's initial conditions are set as: 
\begin{eqnarray*}
\m x(0) &=& \bmat{
	0.4233 &
    377.3780 &
	1.0277 &
	0.1190}^{\top} ,\\ \hat{\m x}(0)&=& \bmat{0&	\omega_0&
	0 &
	0}^{\top},
\end{eqnarray*}
where $\m x(0)$ is the post-contingency system state. The state-space matrices $\mA,\,\mB,$ and $\mC$ in~\eqref{equ:plantDynamics} and~\eqref{equ:plantDynamics2} are omitted for brevity. 
The unknown input coefficients $\mD$ in \eqref{equ:plantDynamics} are randomly chosen and the unknown input vector $\m d(t)=[0.01 \cos(t) \; 0.01 \cos(t) \; 0.01 \cos(t)]^{\top}$ (we consider three unknown inputs with the same magnitude). 

%
To test the robustness of CKF and the observer, a time-varying attack-vector $\m v(t)$ against the four measurements is added to the measured output:
\begin{align}
v_1(t) &= 0.2\sin(t)\nonumber\\
v_2(t) &=  \left\{
    \begin{array}{ll}
      0.5\sin(2t) + \frac{5}{t^2}, & \quad 5<t<6 \\ 
      0, & \quad \textrm{otherwise} \\
    \end{array}
  \right.
\label{equ:attackvector}\\
v_3(t) &= 0.3\cos(t)\nonumber\\
v_4(t) &= 0.2\nonumber.
\end{align}
This attack vector is manually chosen, showing different scenarios when the measured signals can be altered without a clear, known disturbance or attack signal (see Remark~\ref{rem:CAs}). 

Gaussian process and measurement noise are added. 
The process noise covariance is a diagonal matrix, whose diagonal entries are the square of 10\% of the largest state changes, as in \cite{zhou}, and therefore
$\m Q=\textrm{diag}(10^{-6} ,10^{-5} ,1.47 \times 10^{-10},1.17 \times 10^{-8}).$
The measurement noise covariance is $\m R=10^{-4}\mI_4$ where $\m I_4$ is an identity matrix with dimension 4.

\subsection{CKF Estimation Results}
\begin{figure}
	\centering
	\includegraphics[scale=0.26]{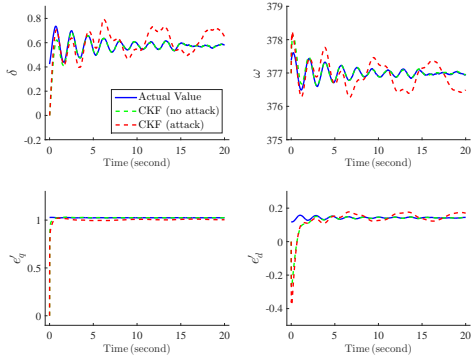}
	\caption{Comparison between the actual states and estimated ones for the SMIB system utilizing CKF without unknown inputs. Green and red dotted lines denote the case without or with cyber attack, respectively.}
	\label{fig:CKF}
\end{figure}

\begin{figure}
	\centering
	\includegraphics[scale=0.26]{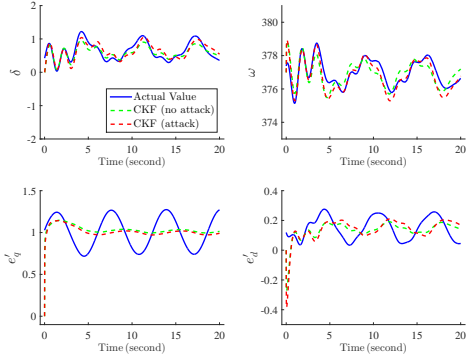}
	\caption{Comparison between the actual states and estimated ones for the SMIB system utilizing CKF with unknown inputs. 
	Green and red dotted lines denote the case without or with cyber attack, respectively.}
	\label{fig:CKF_unknowninput}
\end{figure}

The initial estimation error covariance matrix is chosen as a diagonal matrix whose diagonal entries are the square of 1\% of the initial estimated states. 
If an initial estimated state is zero, the corresponding entry is set to be $0.01^2$. Therefore, $\m P_0=\textrm{diag}(10^{-4},14.2122,10^{-4},10^{-4}).$
The estimation of the states is shown in Figs. \ref{fig:CKF} and \ref{fig:CKF_unknowninput}. It is seen that in the absence of attack-vectors and unknown inputs, 
the estimates from CKF can converge to real states. However, when there are unknown inputs and/or cyber attack, CKF-based estimator fails to track the real states, indicating that CKF is not good at dealing with unknown inputs or cyber attack although it can filter noise.  


\begin{figure}
 	\centering
 	\includegraphics[scale=0.26]{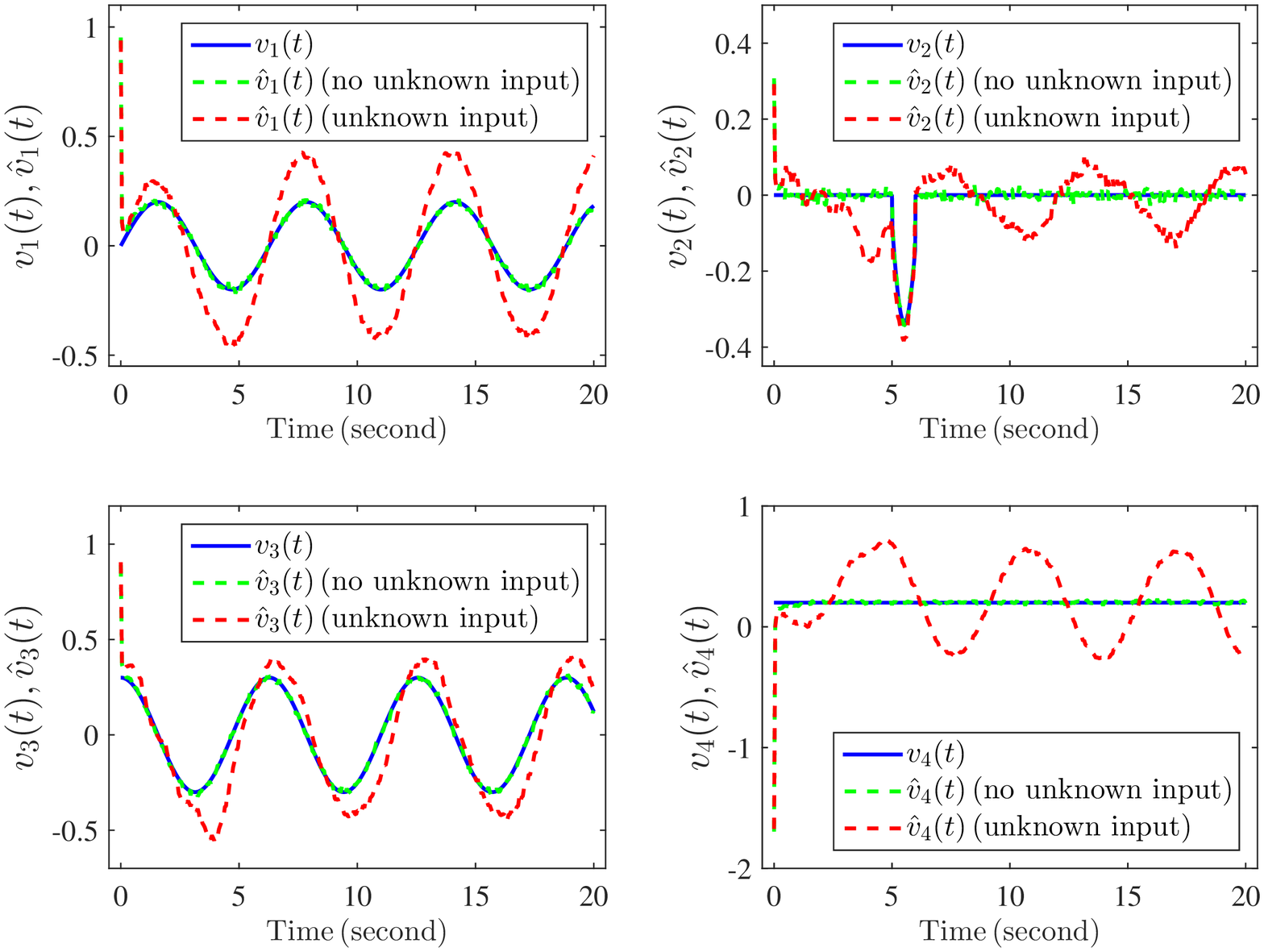}
 	\caption{Attack-vector estimation for the SMIB system utilizing CKF. Green and red dotted lines denote the case with or without unknown inputs, respectively. Attack-vector is added for the two cases.}
 	\label{fig:CKF_AV}
 \end{figure}

The estimation results of attack vectors for the cases with or without unknown inputs are shown in Fig. \ref{fig:CKF_AV}. 
The estimated attach vector is the difference between the estimated output, $\hat{\m y}(t)$, and the actual, measured output, $\m y(t)$.
It is seen that the estimated attack vectors except the second one can converge to real attack vectors under no unknown inputs. 
However, when there do exist unknown inputs, the estimation of attack vectors is poor. 
This is because the CKF is not designed to inherently deal with unknown input vectors with unknown distributions. 

\subsection{Results for Observers}

The observer presented in Section~\ref{sec:TwoObserversA} is presented conceptually in Section~\ref{sec:remarks}. In this section, we briefly present the solution of the observer gain in Section~\ref{sec:TwoObserversB}. The Lipschitz parameters is set to unity, i.e., $\rho=\varphi=\mu=1$.  Using CVX, a package for solving convex optimization problems~\cite{CVX1} on MATLAB, variables $\epsilon_1, \epsilon_2$ and $\sigma$ and a symmetric positive semi-definite matrix $\mP$, are computed:
\EQ
\label{P}
\qquad\;\; \epsilon_1=0.0122, \epsilon_2=0.0144, \sigma=6.424, \\
\small \m P=\MT 	0.4894 & -0.0178 & 0.0627 & -0.4632 \\ 
-0.0178 & 0.0053 & -0.0007 & 0.0067 \\ 
0.0627 & -0.0007 & 0.7708 & 0.0274 \\ 
-0.4632 & 0.0067 & 0.0274 & 0.4945   \EM. \notag
\EE
Then, the observer gain-matrix $\mL$ is computed: 
\EQ
\label{L}
\small
\mL=\frac{\sigma}{2}\mP^{-1}\mC^{\top}= 
\MT -6.023 & 15.9308 & 31.8673 & 12.0481 \\ 
-15.7483 & 42.5073 & 85.0299 & 31.5023 \\ 
4.2075 & 0.063 & 0.1261 & -8.4166 \\ 
-3.1177 & 8.6955 & 17.3942 & 6.2365   \EM, \notag
\EE
a feasible solution for the linear matrix inequality~\eqref{equ:LMIObserverDes}.

\begin{figure}
	\centering
	\includegraphics[scale=0.26]{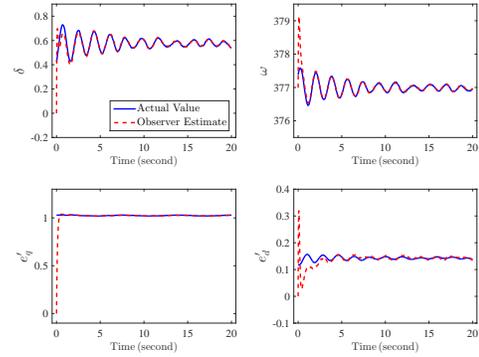}
	\caption{Comparison between the actual states and estimated ones for the SMIB system utilizing the designed observer.} 
	\label{fig:Obsvs}
\end{figure}

\begin{figure}
	\centering
	\includegraphics[scale=0.26]{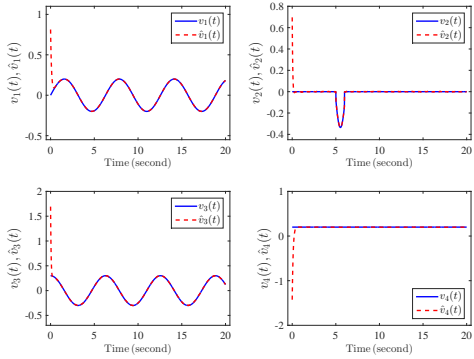}
	\caption{Attack-vector estimation for the SMIB system utilizing the observer under unknown inputs and attack-vectors.}
	\label{fig:ObsvsRealAV}
\end{figure} 

 The estimation of the states using the designed observer is shown in Fig.~\ref{fig:Obsvs}. 
 The observer succeeds in tracking the actual state under the presence of output  attack vector 
 and unknown inputs. 
 The designed observer does not necessitate the knowledge of the initial conditions of the plant. 
 In fact, other numerical simulations show that any observer initial conditions would still lead to accurate state estimation.
 
 Fig.~\ref{fig:ObsvsRealAV} shows the estimation of the predefined attack vector, $\m v(t)$. 
 As shown in Fig.~\ref{fig:ObsvsRealAV}, the attack-vector estimator spectacularly accomplishes near-perfect disturbance estimation. Note that even though $\m v_2(t)$ is discontinuous, the estimator perfectly tracks the attack-vector value. 
\begin{rem}
	The attack-vector estimator can be utilized to identify the type of attack vectors and measurements' disturbances against PMU outputs. Once estimated, the effect of cyber-attacks can be nullified by designing inputs that negate the effect of attack-vectors on state estimation. Furthermore, PMU channels experiencing high output disturbances or attacks might be diagnosed for potential malfunctioning or attacks, as illustrated by NESCOR failure scenarios (see Remark \ref{rem:CAs}). 
\end{rem}
%
%
%
%
%
%
%
%
\section{Remarks and Insights}~\label{sec:remarks}
Various functionalities of DSE methods and their strengths and weaknesses relative to each functionality are presented in this section based on (a) the technical, theoretical capabilities and (b) experimental results of the aforementioned techniques. 
Table~\ref{tab:ComparisonTable} represents a summary of this section. 
\begin{table*}[t]
    \footnotesize
    \renewcommand{\arraystretch}{1.1}
	\centering
	\caption{Comparison between DSE techniques' functionalities and characteristics (`---' indicates a lack of yes/no answer)}
	\begin{tabular}{lcccccr}
		\toprule
		\multirow{2}{*}{\textbf{Functionality/Characteristic}} & \multicolumn{3}{c}{\textbf{Kalman Filter Derivatives}} & \multicolumn{2}{c}{\textbf{Robust Observers}} &  \\
		\cline{2-6}
		& \multicolumn{1}{c}{EKF} & \multicolumn{1}{c}{UKF} & \multicolumn{1}{c}{CKF (Sec.~\ref{sec:CKFF})} & \multicolumn{1}{c}{Obs. 1 
		(Sec.~\ref{sec:TwoObserversA})} & \multicolumn{1}{c}{Obs. 2 (Sec.~\ref{sec:TwoObserversB})} &  \\
		\hline
		\textit{System's Nonlinearities} &    \xmark  & \cmark      &  \cmark      &  \cmark     &  \cmark   &  \\
		\textit{Feasibility} &    ---     &      ---   &   ---      &   \xmark     & \xmark      & \\
		\textit{Tolerance to Different Initial Conditions} &   \xmark    &   \xmark    &  \xmark     &  \cmark       &   \cmark      &  \\
		\textit{Tolerance to Unknown Inputs} & \xmark      & \xmark      & \xmark      &   \cmark    & \cmark      &  \\
		\textit{Tolerance to Cyber-Attacks} & \xmark      & \xmark      & \xmark      &   \xmark    & \cmark      &  \\
		\textit{Tolerance to Process and Measurement Noise} &    \cmark  & \cmark      &  \cmark      &  \cmark     &  \cmark   &  \\
		\textit{Convergence} & ---      & ---    & ---     &   \cmark    & \cmark      & \\
		\textit{Numerical Stability} & ---      & ---    & ---     &   \cmark    & \cmark      & \\
		\textit{Computational Complexity} &  $\mathcal{O}(n^3)$  & $\mathcal{O}(n^3)$  & $\mathcal{O}(n^3)$   & $\mathcal{O}(n^3)$    &  $\mathcal{O}(n^3)$      &  \\
		\bottomrule
	\end{tabular}%
	\label{tab:ComparisonTable}%
\end{table*}%
\begin{itemize}\setlength\itemsep{0.2em}
	\item \textit{Nonlinearities in the Dynamics:} The developed estimation methods in this paper all operate on the nonlinear dynamics 
	unlike the EKF that, by design, assumes a linearized 
	version of the system dynamics. The presented observers in Section~\ref{sec:TwoObservers} assume, nonetheless, that the nonlinear component of the system dynamics is bounded, which is not a restricting assumption for most power systems. 
	
	\item \textit{Solution Feasibility:} The main principle governing the design of most observers is based on finding a matrix gain(s) satisfying a certain condition --- a solution to a matrix inequality or to a \textit{feasibility problem.} 
	Moreover, the gain-matrices for the two observers in Section~\ref{sec:TwoObservers} are solved via formulating a feasibility problem. The state estimates are guaranteed to converge to the actual ones if a solution to the LMI exists. In contrast, KF methods do not require that.
	
	\item \textit{Unknown Initial Conditions:} Unlike Kalman filters, most observer designs are independent on the knowledge of the initial conditions of the system: if the estimator's initial condition is chosen to be reasonably different from the unknown, actual one, estimates generated from KF methods might not converge to the actual ones. For example, if the initial estimated state is chosen as 
	$\hat{\m x}'(0)= \bmat{3 &	\omega_0 & 
		0 &
		0}^{\top}$, the CKF, UKF, and EKF estimates diverge.
	
	\item \textit{Dynamics Subject to Unknown Inputs:} As illustrated in the previous section, CKF-generated estimates is inferior in comparison with the estimation from the presented observer. Many derivatives of the KF are not designed to deal with varying unknown inputs, as the ones assumed in the paper. Similar conclusion applies for other KF methods. 
	
	\item \textit{PMU Measurements Subject to Cyber-Attacks:} The observer for the nonlinear system outperforms the CKF, UKF, and EKF in the state estimation under attack-vectors as well as the estimation of the attack-vector.
	
	\item \textit{Tolerance to Process and Measurement Noise:} The observers under consideration are tolerant to measurement and process noise similar to the ones assumed for KFs. By design, KF techniques are developed to deal with such noise. 
	
	\item \textit{Estimation Convergence Guarantees:} Observers provide theoretical guarantees for the estimation convergence. However, for KF there is no strict proof to guarantee that the estimation converges to actual states.
	
	\item  \textit{Numerical Stability:} Observers do not have numerical stability problems while UKF can encounter numerical instability because the estimation error covariance matrix is not always guaranteed to be positive semi-definite.
	
	\item \textit{Tolerance to Parametric Uncertainty:} KF-based methods can tolerate inaccurate parameters to some extent. Dynamic observers deal with parametric uncertainty in the sense that all uncertainties can be augmented to the unknown input component in the state dynamics ($\mD\m d(t)$). After redefining $\mD\m d(t)$ due to uncertainty, the observer is redesigned and the observer gain is updated to reflect the changes. 

	\item \textit{Computational Complexity:} The observers presented in the paper, CKF, UKF, and EKF all have computational complexity of $\mathcal{O}(n^3)$ \cite{van2001square,CKF}. However, observers require fewer computational steps. Since the observers' matrix gains are obtained offline by solving LMIs, observers are easier to implement as only the dynamics are needed in the simulations (e.g., Equations~\eqref{equ:plantDynamics}-\eqref{equ:plantDynamics2} and~\eqref{equ:ObserverTwo}). Table~\ref{tab:RT} shows a comparison of the running time of CKF, UKF, and the observer from Section~\ref{sec:TwoObserversB} for the SMIB system --- the estimation with the observer being the fastest among the three methods. 
\end{itemize}
\begin{table}[h]
    \footnotesize
    \renewcommand{\arraystretch}{1.3}
	\centering
	\caption{Running Time of the Three Estimators for SMIB System}
	\begin{tabular}{cccc}
		\toprule
		\multirow{2}{*}{Initial Conditions}  &  \multicolumn{3}{c}{Running Time (seconds)} \\ \cline{2-4}
	 & \multicolumn{1}{c}{CKF} & \multicolumn{1}{c}{UKF} & \multicolumn{1}{c}{Obs.2 (Sec.~\ref{sec:TwoObserversB})} \\
	 \midrule 	 
		$\hat{\m x}(0)=[0\,\,  \omega_0\,\,   0\,\,   0]^{\top}  $ & 3.28  & 6.30   & 0.76 \\
		$\hat{\m x}^{\prime}(0)=[3\,\,  \omega_0\,\,   0\,\,   0]^{\top}  $ & 3.27  & 6.27  & 0.76 \\
		\bottomrule
	\end{tabular}%
	\label{tab:RT}%
\end{table}%

\section{Conclusion}~\label{sec:conc}

In this paper, we discuss different dynamic state estimation methods 
by presenting an overview of state-of-the-art estimation techniques and developing alternatives, including the cubature Kalman filter and dynamic observers, to address major limitations of existing methods. The proposed methods are examined on a single-machine infinite-bus benchmark system, under unknown inputs and attack-vectors. 
Based on both the technical, theoretical capabilities and the experimental results, 
we summarize the strengths and weaknesses of different techniques, and also provide the reader with recommendations for optimal DSE in power networks. 
\bibliographystyle{ieeetran}
\bibliography{main}
\end{document}